\documentclass{aastex63}
\usepackage{epsfig}
\usepackage{float}

\begin{document}

\title{Limiting the abundance of LIGO/Virgo black holes with microlensing \\ observations of quasars of finite size}

\author{A. ESTEBAN-GUTI\'ERREZ} 
\affiliation{Instituto de Astrof\'{\i}sica de Canarias, V\'{\i}a L\'actea S/N, La Laguna 38205, Tenerife, Spain}
\affiliation{Departamento de Astrof\'{\i}sica, Universidad de la Laguna, La Laguna 38206, Tenerife, Spain}
\author{E. MEDIAVILLA}
\affiliation{Instituto de Astrof\'{\i}sica de Canarias, V\'{\i}a L\'actea S/N, La Laguna 38205, Tenerife, Spain}
\affiliation{Departamento de Astrof\'{\i}sica, Universidad de la Laguna, La Laguna 38206, Tenerife, Spain}
\author{J. JIM\'ENEZ-VICENTE}
\affiliation{Departamento de F\'{\i}sica Te\'orica y del Cosmos, Universidad de Granada, Campus de Fuentenueva, 18071 Granada, Spain}
\affiliation{Instituto Carlos I de F\'{\i}sica Te\'orica y Computacional, Universidad de Granada, 18071 Granada, Spain}
\author{N. AG\"UES-PASZKOWSKY}
\affiliation{Instituto de Astrof\'{\i}sica de Canarias, V\'{\i}a L\'actea S/N, La Laguna 38205, Tenerife, Spain}
\affiliation{Departamento de Astrof\'{\i}sica, Universidad de la Laguna, La Laguna 38206, Tenerife, Spain}
\author{J. A. MU\~NOZ}
\affiliation{Departamento de Astronom\'{\i}a y Astrof\'{\i}sica, Universidad de Valencia, 46100 Burjassot, Valencia, Spain.}
\affiliation{Observatorio Astron\'omico, Universidad de Valencia, E-46980 Paterna, Valencia, Spain}
\author{S. HEYDENREICH}
\affiliation{Argelander-Institut f\"ur Astronomie, Auf dem H\"ugel 71, 53121, Bonn, Germany}

\begin{abstract}

  We present a simple but general argument that strongly limits the abundance of Primordial Black Holes (PBHs) (or other unknown population of compact objects) with masses similar to those determined by LIGO/Virgo from BH binary mergers. We show that quasar microlensing can be very sensitive to the mass of the lenses, and that it is able to distinguish between stars and BHs of high mass, when the finite size of the source is taken into account. A significant presence of massive BHs would produce frequent high flux magnifications (except for unrealistically large sources) which have been very rarely observed. On the contrary, a typical stellar population would induce flux magnifications consistent with the observations. This result excludes PBHs (or any type of compact object) in the mass range determined by LIGO/Virgo as the main dark matter constituents in the lens galaxies.

\end{abstract}


\section{Introduction \label{intro}}

 The discovery of gravitational waves from binary black hole (BBH) mergers by the LIGO/Virgo collaboration (cf{\color{blue}.} GWTC-1 and GWTC-2 by Abbott et al. 2019a and Abbott et al. 2020a) with masses higher than previously expected for BHs of stellar origin (but also the low effective spins of the components) renewed in the last years the interest on the possibility that some of these BHs were of primordial origin, and even that these primordial black holes of intermediate mass (20$-$200 $M_\odot$), not excluded by galactic microlensing, could constitute a significant fraction of the dark matter in the Universe (Carr \& K\"uhnel, 2020).
\\

Quasar microlensing (Chang \& Refsdal 1979; Wambsganss 2006) provides an alternative path to study the abundance, not only of BHs, but also of any type of compact object. The description of this phenomenon is simple{\color{blue}:} an intervening galaxy (the lens) deflects the light from a distant quasar forming several images. These images are usually seen through the lens galaxy and, as far as the matter distribution in the galaxy is not smooth but granulated in compact objects (stars, BHs, etc.), the light beams can suffer new secondary deflections producing several microimages (that can not be resolved by telescopes). The primary observational effect of this image splitting is a change in the flux of the images (microlensing flux magnification). The relevant question we will address here is: Can the amplitude and frequency of these microlensing flux magnifications inform us about the mass of the microlenses? Or, more specifically: can microlensing observations unequivocally reveal the presence of LIGO/Virgo BHs?
\\

Previous studies based on quasar microlensing (Mediavilla et al. 2017, Esteban-Guti\'errez  et al. 2020) do not support the existence of a significant population of intermediate mass BH. However, neither these negative results from microlensing, nor the existence of new proposed paths to explain the stellar formation of BH of intermediate mass (see, e.g., Abbott et al. 2020b), seem to have had a major impact in stopping the speculation about the existence of  a population of PBHs that could account for the dark matter. This fact may be related to the indirect approach and rather complex statistical modelling involved in those microlensing studies which we intend to avoid here.
\\

As we will discuss, the key parameter that determines the differences in amplitude and frequency of microlensing magnifications corresponding to microlenses populations of different masses is the size of the lensed object, which in our case is the size of the quasar accretion disk at (rest frame) UV wavelengths.
The motivation of this work is, then, to show using broadly applicable arguments, that quasar microlensing is very sensitive to the mass of BHs in the range of masses of LIGO/Virgo detections. Specifically, we will show that there is a strong difference in predicted microlensing magnifications for stars and LIGO/Virgo BHs when a source of finite size is considered. We leave a thorough statistical analysis with quantitative estimates of the limits in the abundance of BHs for an accompanying paper.
\\

The letter is organized as follows. In \S2 we analyze the impact of the size of the source in the probability distribution of observing a given microlensing magnification and compare the results with available microlensing observations. In \S3 we discuss the differences in microlensing of quasars by stars or by LIGO/Virgo BHs. Finally, in this same section (\S3), we summarize the conclusions.
\\

\section{Results \label{results}}

\begin{figure*}[ht]
  \includegraphics[scale=0.68]{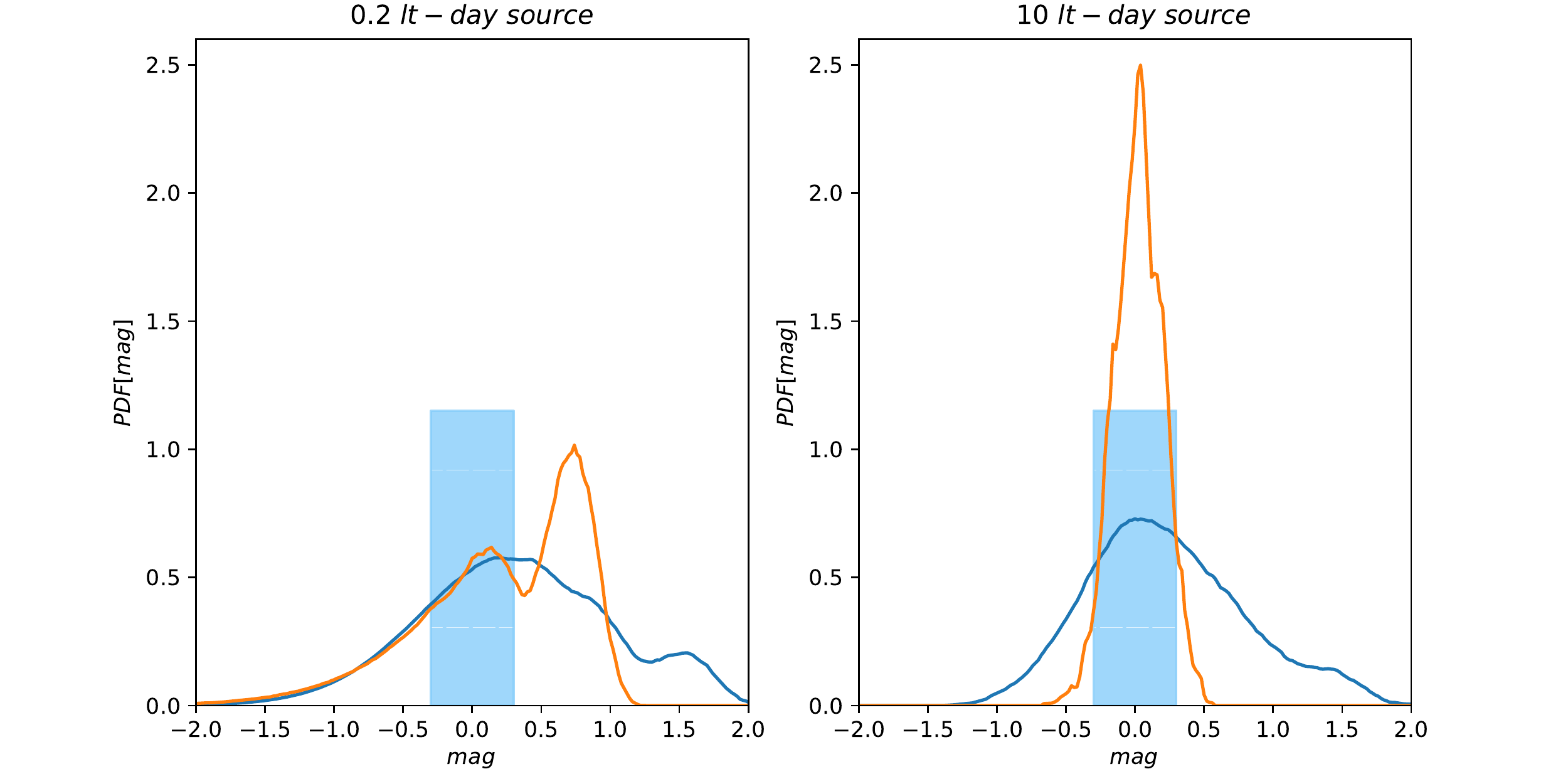}
    \caption{Probability distributions of microlensing flux magnifications for a population of 20\% of stars of $0.2 M_\odot$ plus 80\% of smooth dark matter (orange curve) and a  population of 20\% of stars of $0.2 M_\odot$ plus 80\% of LIGO/Virgo type BHs of  $30 M_\odot$ (blue curve). Left panel for a source of 0.2 light days. Right panel for a source of 10 light days. Scaling has been chosen to give unit area under the PDF curves. The shadowed blue marks the region containing $\sim$70\% of observed microlensing magnifications (Pooley et al. 2007) i.e., with the ordinate chosen to enclose 70\% of the probability (see text).}  \label{fig}

\end{figure*}

We want to study the impact of the mass of the microlenses in both the amplitude and frequency of magnifications, focusing on to what extent a population of LIGO/Virgo type BHs can account for the unrevealed dark matter in the lens galaxies. To do that, we are going to calculate the probability of measuring a given microlensing flux magnification for two different populations of microlenses, both with a 20\% of matter in stars (with $m=0.2M_\odot$, typical of old stellar populations), but one with the remaining 80\% of matter in smooth dark matter\footnote{This is the typical baseline scenario (Schechter et al. 2014, Jim\'enez-Vicente et al. 2015a, 2015b) in which the population of microlenses contributes to 20\% to the total mass density (including dark matter), and the remaining 80\% is in the form of a smooth distribution of matter.}, and the other in typical LIGO/Virgo BHs with $m=30M_\odot$. Following the standard procedure, we obtain the Probability Density Function (PDF) of the flux magnification from the histogram of the magnification maps\footnote{Obtained by tracing light rays backwards from the image to the source through the random distribution of microlenses (Kayser et al., 1986)}. In these calculations, we have considered a typical lens system\footnote{We take $\kappa=\gamma=0.45$, $z_l=0.5$, and $z_s=2$.} with a mean flux magnification of 10. We use a pixel size of 0.2 light days for the magnification maps, very much smaller than typical quasar sizes. The PDFs of measuring a certain flux magnification for both mass distributions are shown in Figure \ref{fig} (left). Both PDFs match very well in the negative magnitude region (corresponding to microlensing flux increase), although they show clear differences in the positive wing (corresponding to flux decrease). \\ 

It is worth wondering, at this point, how well do these probability distributions compare to actual observations. The optical fluxes (normalized to their expected model values) reported by Pooley et al. (2007) (their Table 5) for nine\footnote{We have eliminated Q2237+0305 from their sample in this comparison, as this system is produced by a nearby lens, and has nearly 100\% of the mass density in form of compact objects.} quadruple lens systems are very well suited for a quick comparison. While according to Pooley et al. (2007) a $\sim$70\% of the observed magnifications fall in the (-0.3,0.3) magnitudes range (region shadowed in blue in Figure \ref{fig}) our simulations for an {\em infinitesimal} source predict only a $\sim$30\% of measurements in this range. As we will see below, this strong discrepancy is originated by the fact that real quasars have sizes much larger than the pixel size used in our calculations.
To take this fact into account, we can calculate the corresponding probability distributions for an extended source. We have chosen here to model the source with a Gaussian brightness profile, but this is known to have little effect on the microlensing magnification probability distributions (Mortonson et al. 2005, Mu{\~n}oz et al. 2016). We consider a source size of 10 light-days, which approximates the Einstein radius (the natural length scale of lensing) of the $m=0.2M_\odot$ stars, yet it is much smaller than the Einstein radius of the BHs ($\sim$112 lt-days). Nevertheless, conclusions remain essentially unchanged for any reasonable source size compatible with observations.
\\

The resulting probability distributions for this larger source are shown in Figure  \ref{fig} (right). In this Figure, it can be clearly seen that while the PDF of the stars becomes significantly narrower (therefore predicting less extreme magnifications in better agreement with observations)
  the PDF of the BHs has changed much less (still predicting over $\sim$60\% of magnifications outside the (-0.3,0.3) magnitude range).
 The reason for this different behavior is indeed rather simple: for a finite source size, the gradients in the magnification maps produced by lenses with an Einstein radius comparable to the source size get blurred/averaged, while this effect is minimized if the source has a negligible size compared to the Einstein radius of the lens. 
 \\
 
 We would like to stress again that although we have chosen here some values of the parameters to illustrate the principle, this result is, nevertheless, very general, and has little dependence on the specific choice as long as it is in reasonable agreement with observations. 
\\

\section{Discussion and Conclusions\label{discussion}}

 An immediate result of the previous section is that we need to consider sizes of quasars comparable to the Einstein radius of the lenses in order to narrow the predicted PDF to approximate (even roughly) the experimental histogram of microlensing magnifications. If BHs were to account for a significant fraction of the mass in the lens galaxies, this would imply source sizes of $\gtrsim 100$ light-days, which is absolutely discarded by reverberation mapping estimates of the size of the accretion disks of quasars (see e.g. Edelson et al. 2015, Fausnaugh et al. 2016, Jiang et al. 2017, Cackett et al. 2018, Homayouni et al. 2019 and Yu et al. 2020). On the contrary, for typical estimated sizes of quasar accretion disks at these wavelengths of a few light days, the PDF predicted by a population of stars is reasonably consistent with the observations. Therefore, if the dark matter of the lens galaxies were in the form of BHs of the masses detected by LIGO/Virgo, much larger microlensing magnifications should have been regularly observed. As this is not the case, an explanation to dark matter based on this kind of objects can safely be discarded.
\\

On the other hand, although we have used here for comparison the estimates of microlensing magnifications from Pooley et al. (2007) (because the reported observed fluxes normalized to model predictions provide directly the flux magnifications), the rarity of high flux magnifications is also extensively confirmed by the statistics of differential flux magnifications between images (Mediavilla et al. 2009, Fian et al. 2016, Fian et al. 2018) and by the microlensing induced variability observed in the light curves of lensed quasar images (see, e.g. Mediavilla et al. 2016 and references therein).
\\

Regarding the possibility of a partial explanation of dark matter in terms of LIGO/Virgo BHs, we have also calculated the PDFs of populations including less than an 80\% of BHs, just to a 10\%, confirming the significant over-prediction of non observed large microlensing magnifications, even for this small abundance of BHs. In an accompanying paper we discuss the likelihood of a population of BHs in all the range of BHs abundances.
\\

From the above analysis, we can reach the following conclusions:
\begin{enumerate}
\item Observed microlensing magnification statistics in optical observations of lensed quasars can only be explained if the source (the quasar accretion disk) size is comparable to the Einstein radius of the microlenses.
\item For reasonable values of the size of the observed quasars, if the dark matter were in form of compact objects with masses in the range of the BHs detected by LIGO/Virgo, much larger magnifications should have been frequently observed. The absence of such observations is therefore a strong proof that the dark matter in lens galaxies is not formed by these objects.
  \end{enumerate}

 The above conclusions are based on very general arguments which do not depend on specific models. Notwithstanding, a quantitative full Bayesian statistical modelling of a mixed populations of stars and BHs setting quantitative limits to the abundance of the latter is included in an accompanying paper.

\acknowledgments{We thank the anonymous referee for ideas and comments, which greatly contribute to enhance the scope of our paper. This research was supported by the Spanish MINECO with the grants AYA2016-79104-C3-1-P and AYA2016-79104-C3-3-P.  J.J.V. is supported by the project AYA2017-84897-P financed by the Spanish Ministerio de Econom\'\i a y Competividad and by the Fondo Europeo de Desarrollo Regional (FEDER), and by project FQM-108 financed by Junta de Andaluc\'\i a. AEG thanks the support from grant FPI-SO from the Spanish Ministry of Economy and
Competitiveness (MINECO) (research project SEV-2015-0548-17-4 and predoctoral
contract BES-2017-082319).}

\end{document}